\documentclass[9pt,journal]{IEEEtran}

\usepackage[noadjust]{cite}
\usepackage{rotating}
\usepackage{diagbox}
\usepackage{colortbl}
\usepackage{tikz}
\newcommand*\circled[1]{\tikz[baseline=(char.base)]{
            \node[shape=circle,draw,inner sep=1pt] (char) {#1};}}

\usepackage{hyperref}
\newcommand{\figref}[2][{}]{\hyperref[#2]{\figurename~\ref{#2}#1}}

\usepackage{amsmath}
\usepackage{amssymb}
\usepackage{cleveref}
\usepackage{booktabs}
\usepackage{tabu}
\usepackage{longtable}
\usepackage{bm}
\usepackage{multirow}

\usepackage{arydshln}

\usepackage{array}

\ifCLASSOPTIONcompsoc
 \usepackage[caption=false,font=normalsize,labelfont=sf,textfont=sf]{subfig}
\else
 \usepackage[caption=false,font=footnotesize]{subfig}
\fi

\usepackage{stfloats}

\ifCLASSOPTIONcaptionsoff
 \usepackage[nomarkers]{endfloat}
\let\MYoriglatexcaption\caption
\renewcommand{\caption}[2][\relax]{\MYoriglatexcaption[#2]{#2}}
\fi

\usepackage{listings}
\usepackage{courier}

\usepackage{url}

\begin{document}
\title{Effective Parallelism for Equation and Jacobian Evaluation in Large-Scale Power Flow Calculation}

\author{Hantao~Cui,~\IEEEmembership{Senior Member,~IEEE,}
        Fangxing~(Fran)~Li,~\IEEEmembership{Fellow,~IEEE,}
        Xin~Fang,~\IEEEmembership{Senior Member,~IEEE}%
\thanks{H. Cui and F. Li are with the Department
of Electrical Engineering and Computer Science, The University of Tennessee, Knoxville,
TN, 37996 USA. E-mail: fli6@utk.edu.}%
\thanks{X. Fang was with the Department of EECS at the University of Tennessee and the CURENT research center, Knoxville, TN, 37996 USA.}
\thanks{This work was supported in part by the Engineering Research Center Program of the National Science Foundation and the Department of Energy under NSF Award Number EEC-1041877 and the CURENT Industry Partnership Program.}%
}%
\markboth{Preprint to be submitted to IEEE Power Engineering Society Letters}%
{Cui \MakeLowercase{\textit{et al.}}: Effective Parallelism for Power Flow Calculation}

\maketitle

\begin{abstract}
This letter investigates parallelism approaches for equation and Jacobian evaluations in large-scale power flow calculation.
Two levels of parallelism are proposed and analyzed: inter-model parallelism,
which evaluates models in parallel,
and intra-model parallelism, which evaluates calculations
within each model in parallel.
Parallelism techniques such as multi-threading and single instruction multiple data (SIMD) vectorization are discussed, implemented, and benchmarked as six calculation workflows.
Case studies on the 70,000-bus synthetic grid show that equation evaluations can be accelerated by ten times, and the overall Newton power flow
advances the state of the art by 20\%.
\end{abstract}
\begin{IEEEkeywords}
Power flow calculation, parallelism, multi-threading, single instruction multiple data (SIMD).
\end{IEEEkeywords}

\IEEEpeerreviewmaketitle

\section{Introduction}
\IEEEPARstart{P}{ower} flow calculation is a fundamental routine
widely used for power system analysis and operation. 
In power systems with a large contingency set and various renewable scenarios,
the calculation speed of power flow is crucial to ensure system security.
Recent CPUs have stalled on clock rate and 
started to pack more cores and provide sophisticated
instruction sets.
Therefore, power flow workflows need to be fine-tuned
to utilize new computing hardware.

Power flow calculation consists of 
four sequential tasks: 
(1) updating equation residuals,
(2) updating Jacobian elements, 
(3) solving sparse linear equations, 
and (4) updating variable values and checking
for convergence.
Generally, task (4) is light-weight, and task (3) is handled by highly efficient external
solvers \cite{6672387}, such as KLU on CPU or CuSparse on GPU \cite{zhou2017gpu, 7725984}.
We investigate the parallelization within tasks (1) and (2).

Available parallel computing techniques include instruction-level, data, and task parallelism.
Instruction-level parallelism utilizes dedicated instructions available on the platform.
Data parallelism distributes the same computing task on large data sets across multiple processors.
Task parallelism dispatches each processor core, using \textit{multi-processing} or \textit{multi-threading},
for one computing task on the same or different data.
Given the size of power flow problems, 
instruction-level and task parallelism are more relevant.

Related work has explored instruction-level parallelism using 
Single Instruction Multiple Data (SIMD) vectorization
for Task (3), namely, factorizing sparse matrices and solving sparse linear equations,
on GPU \cite{jalili-marandi_simd-based_2010,vilacha_massive_2011}.
CPU multi-processing is reported in \cite{dzafic_real-time_2010,ahmadi_parallel_2018} for power flow,
both using OpenMP for message passing.
A CPU-GPU architecture is proposed in \cite{roberge_parallel_2017} that combines GPU SIMD with multi-processing
for power flow runs of many instances of the same power network. 
To our best knowledge, this letter is the first to quantify the effectiveness of
SIMD and multi-threading for parallelizing the equation and Jacobian evaluations on modern CPUs.

In most existing tools, such as PSAT \cite{milano2005psat} and 
ANDES \cite{cui2020hybrid}, the numerical equations and Jacobian elements
are obtained by calling power flow models in serial,
which is partially due to the limited parallelism supports
in MATLAB and Python.
In modern programming environments, power flow routines with instruction-level and task parallelism 
can be prototyped to provide directions for improving existing tools.

This work studies the parallelism for the equation
and Jacobian evaluation tasks in power flow calculation.
Since these two tasks share the same type of computing demand,
namely, arithmetic evaluation of equations for residuals and Jacobian elements, the parallel procedures are the same.
The main contributions are:
\begin{enumerate}
    \item The concepts of inter- and intra-model
    parallelism are proposed. Implementation,
    advantages and limitations are discussed.
    \item Six workflows, as the combinations of
    two inter-model and three intra-model workflows,
    are benchmarked using large-scale synthetic grids
    with up to 70,000 buses \cite{7725528}.
\end{enumerate}

An an outcome, the Newton power flow package that implements the 
most effective parallelism is over 20\% faster than the state-of-the-art
for large-scale systems,
as will be discussed in \Cref{sec:case-studies}.

\section{Power Flow Models and Workflows}
\subsection{Power Flow Formulation and Models}
This work employs an extended power flow formulation that
models the voltage control of PV generators and 
the voltage and phase angle controls of the slack generator.
In addition the bus voltages and phase angles solved from the traditional formulation, this formulation allows checking reactive power limits in iterations, 
and same formulation can be used before and after 
a PV generator switches to a PQ load.
The formulation is given by the compact notation that
\begin{equation}
    \bm{g}(\bm{y}) = \bm{0} \, , 
\end{equation}
where algebraic equations $\bm{g}$ in \eqref{eq:g_compact}
are organized by grouping active
power mismatches, reactive power mismatches, voltage
control errors, and angle control error in order.
Variables $\bm{y}$ in \eqref{eq:y_compact} are 
grouped by bus voltage angles,
bus voltage magnitude, PV reactive power outputs,
and Slack active power output in order.
\begin{equation}
\label{eq:g_compact}
    \bm{g}(\bm{y}) = 
    \begin{bmatrix}
    \bm{g_p}, \bm{g_q}, \bm{g_V}, {g_\theta} 
    \end{bmatrix}^T \, ,
\end{equation}
\begin{equation}
\label{eq:y_compact}
    \bm{y} = 
    \begin{bmatrix}
    \bm{\theta}, \bm{V}, \bm{Q_g}, {P_s} 
    \end{bmatrix}^T \, .
\end{equation}

This formulation can be extended, for example, to 
include control modes of converters
in power flow.
We also note that this formulation has more equations
and variables than the well-known one.

Like most power flow programs, this work considers
models including bus, PQ load, PV generator,
slack generator, lines (including
the $\pi$-model transmission line
and two-winding transformer) and shunt capacitor.
In particular, line injections are calculated on
a per-line basis and summed at the connected buses.
Since the power injections at each terminal are calculated independently, parallelization can
be possible, as will be discussed in the following.

\begin{figure}
\centering
\includegraphics[width=\columnwidth]{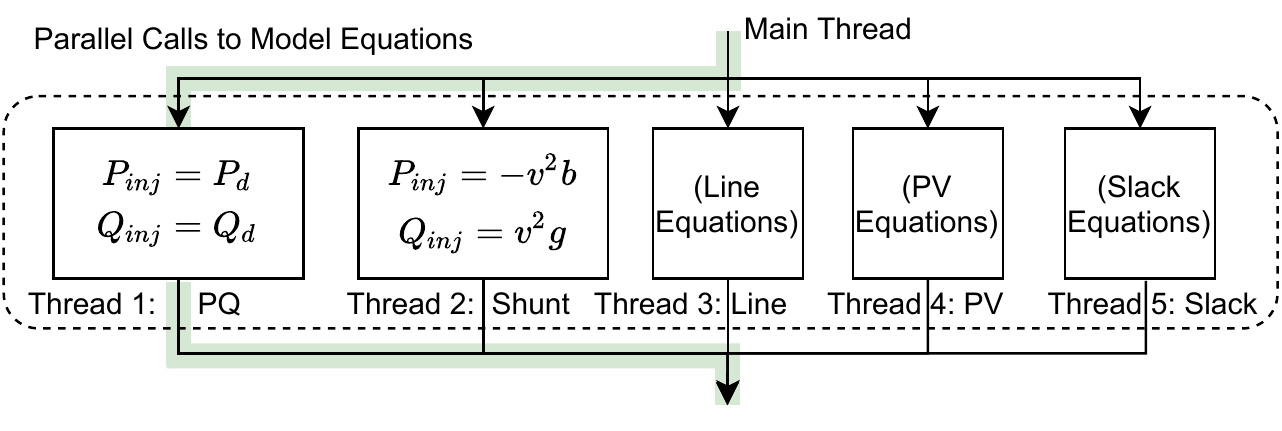}
\caption{Detailed view of inter-model parallelism for equations evaluations.}
\label{fig:inter-model-parallel}
\end{figure}

\subsection{Parallel Workflows: Inter-model and Intra-model Parallelism}
\label{sec:parallel-workflow-intro}

\begin{figure*}
\centering
\includegraphics[width=1.8\columnwidth]{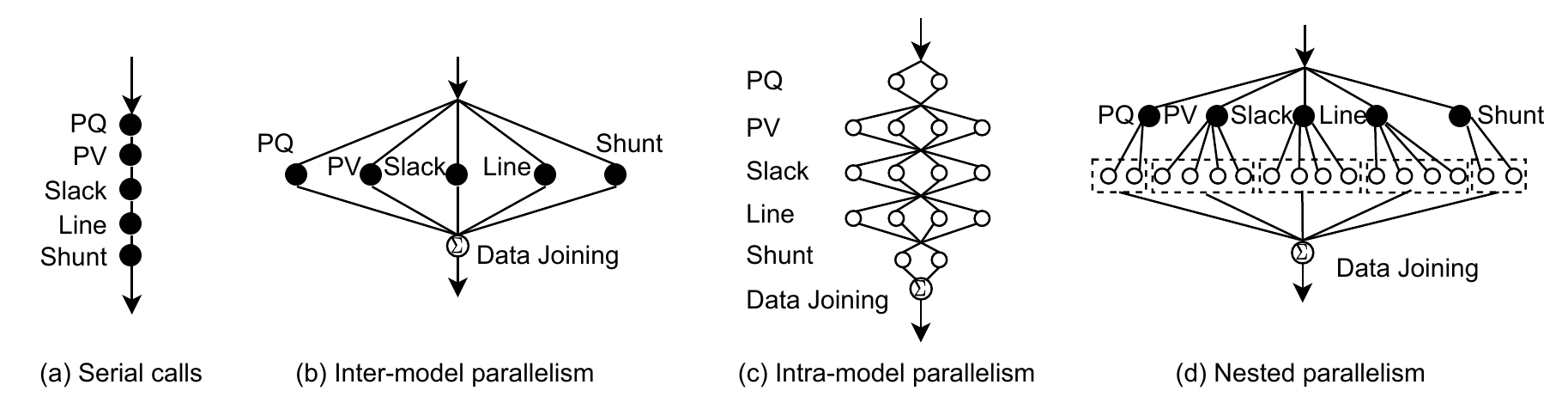}
\caption{Power flow workflows. Solid dots denote the aggregation of all equation/Jacobian evaluations in a model. Circles denote each equation or Jacobian element evaluation.}
\label{fig:parallel-workflows}
\end{figure*}

Two levels of parallelism are proposed for tasks (1) and (2). 
The first level is the \textit{inter-model parallelism},
which aims to evaluate the equations and Jacobian elements of
multiple models in parallel.
A high-level view of inter-model parallelism 
is shown in \figref{fig:parallel-workflows}(b), where each solid dot
represents a model.
A detailed view of the parallel part
is shown in \figref{fig:inter-model-parallel},
where PQ, Shunt, and other models compute their power injection equations
in parallel.
Results from the parallel evaluations
need to be joined as the whole system.
For example, injections from lines, loads, and
generators on the same bus will be joined through summation.

The inter-model parallelism is a coarse-grained approach 
typically assigns one processor core for each function of each model.
However, this workflow has a limitation relevant for
power flow:
all jobs must have been completed before results can be joined.
This limitation is known as Cannikin's Law,
which suggests that, although threads 
run in parallel, the total time is bottlenecked
by the slowest thread.
For large systems, the line model is the slowest
due to the instance number and calculation complexity.

The second level is the \textit{intra-model parallelism},
a fine-grained approach that parallelizes
independent calculations within models.
Consider the active power injections at transmission line terminals:
\begin{equation}
\label{eq:line-inj}
\begin{matrix}
    P_h = v_h^2(g_L + g_{L,h}) - v_h v_k (g_L cos\theta_{hk} + b_L sin\theta_{hk}) \\
    P_k = v_k^2(g_L + g_{L,k}) - v_h v_k (g_L cos\theta_{hk} - b_L sin\theta_{hk}) \, ,
\end{matrix}
\end{equation}
where $v_h$ and $v_k$ are the bus voltages at terminals $h$ and $k$,
and $\theta_{hk}$ is the voltage angle difference,
$g_L$ and $b_L$ are the conductance and susceptance of
the series component, and $g_{L,h}$ and $g_{L, k}$ 
are the conductance of the shunt-component.  
The intra-model parallelism aims to compute $P_h$ and $P_k$
in parallel due to independence.
Two approaches are available for implementing intra-model parallelism:
task parallelism or SIMD vectorization,
which will be discussed in
\Cref{sec:simd}.
Intra-model parallelism can efficiently utilize processors
since independent calculations of the same model 
usually have similar complexity, as are 
$P_h$ and $P_k$ in \eqref{eq:line-inj}.
However, intra-model parallelism is more difficult to 
implement, as one needs to carefully eliminate data racing
within each model.

Further, one can create \textit{nested parallelism}
by combining the inter-model and intra-model
parallelism to assign one core for each model and assign
one core (or uses SIMD) for each equation-level calculation.
At first glance, the idea may be appealing, given that all equation-level
computations are executed in parallel.
However, if task parallelism is used for both inter- and intra-model
parallelism, processor resources may run out quickly. 

\subsection{Task Parallelism: Multi-processing or Multi-threading}
As mentioned, task parallelism can be implemented with multi-processing or multi-threading to utilize multiple cores.
By design, a process is a program in execution composed of code, current activity, data, heap, and stack.
Processes can only share resources through techniques such as message passing or shared memory, which require explicit arrangements by the programmer \cite{silberschatz:2012}.
Therefore, Multi-processing is ideal for computations where data exchanges between individual runs are not required, such as contingency screening and stochastic time-domain simulation \cite{6547228}.

A thread is a basic unit of CPU utilization within a process. By default, a thread shares the resources (such as code and data) of the process to which it belongs.
Empirically, threads consume significantly less time to create and manage \cite{silberschatz:2012}.
Such characteristics make multi-threading suitable for parallelizing equations and Jacobian update functions, which need to be called multiple times and require the passing of inputs (from the previous iteration) and outputs.

\subsection{Single Instruction Multiple Data (SIMD) Vectorization}
\label{sec:simd}
SIMD is a mechanism that vectorizes arithmetic operations
at the processor level.
This vectorization is completely different from MATLAB's
vectorization, which expresses loop operations using vectors.
For example, on Intel processors that supports AVX512,
the 512-bit Advanced Vector Extension instructions,
programs can pack 32 double-precision floating point
calculations per clock cycle.
To compute a power system with 32,000 line devices, for example, 
each equation in \eqref{eq:line-inj} only needs
1,000 evaluations with AVX512.

However, SIMD is not automatic and requires meticulous implementation.
Specifically, the following requirements must be satisfied:
\begin{enumerate}
    \item Use vectors to store numericals in contiguous memory.
    \item Eliminate bound checking when accessing vectors.
    \item
    Use proper directives to suggest SIMD to the compiler.
    \item Inspect the generated machine code to verify vectorizations.
\end{enumerate}

Although SIMD is intricate to implement, the speedup is huge compared with scalar calculations, as will be shown in \Cref{sec:case-studies}.

\section{Valid Implementation and Benchmark}

To utilize multi-core processors, the power flow program must be implemented in a language that (1) compiles into bare-metal machine code (as opposed to virtual machine byte-code), and (2) is capable of dispatching multiple cores.
Unfortunately, none of the two popular languages for scientific computing, namely, MATLAB and Python, meets the requirement.
Due to the global interpreter lock (GIL), Python threads are executed one after another, defeating the purpose of simultaneous multi-threading.
Python's multiprocessing does allow side-stepping GIL to achieve process-based parallelism. However, as discussed, processes are computationally costly to start and require careful arrangements by the programmer for data sharing.
MATLAB provides a multi-processing-based parallel for-loop (parfor) solution, which occurs only if the parfor-loops are compiled into MEX functions using an OpenMP-compliant compiler.
The restriction renders MATLAB unfriendly to structure parallel programs.

This work implements the multi-threaded workflows in the Julia language \cite{bezanson2017julia},
which is high-level, just-in-time compiled, and multi-threaded.
Julia's high-level syntax allows quick prototyping of the power flow models and equations. 
Under the hood, Julia uses the Low-Level Virtual Machine (LLVM) compiler to translate the code into platform-specific, highly optimized machine code.
Most importantly, Julia can readily dispatch processor cores for multi-threading. 

Run-time memory allocation is another factor that invalidates multi-threading efforts.
One of the core challenges in high-performance computing is to minimize memory access time. 
Run-time allocations will incur access time and trigger garbage collection, which is another performance deal-breaker.
In this work, equation calls are allocation-free, and Jacobian calls only allocate while assembling triplets into sparse matrices.

\section{Case Studies}
\label{sec:case-studies}

\subsection{Test Systems and Environments}
The effectiveness of parallelism is best validated using large test systems on recent-generation hardware.
The study utilizes the Synthetic 70,000-bus system, which contains 88,207 branches and transformers, 160,780 variables, and 998,495 non-zero Jacobian elements.
Simulations are performed on a workstation with the six-core Intel Xeon W-2133 and 32 GiB of memory running Ubuntu 16.04, Julia 1.5.2, Python 3.8.5, NumPy 1.19.1, and MATLAB 2020a.
Computation time is measured using the Newton power flow.

\subsection{The Effectiveness of SIMD Vectorization}
SIMD is first evaluated for effectiveness by comparing implementations in Julia, Python, and MATLAB.
The times to calculate the four equations of Line (active and reactive power injections at two terminals) are reported.
SIMD is verified in the generated assembly code with instructions such as \lstinline{vaddpd} and \lstinline{vmulpd}, where \lstinline{v} is vectorized, \lstinline{add} and \lstinline{mul} are addition and multiplication, and \lstinline{pd} indicates packed data (as opposed to scalar data \lstinline{sd}).

As reported in \Cref{tab:time-equations-simd}, 
turning on SIMD for the same Julia code makes the computation is an order of magnitude faster.
MATLAB records satisfactory performance compared with the SIMD version in Julia, because MATLAB utilizes SIMD under the hood for array operations.
On the other hand, NumPy is relatively slow due to its limited SIMD support, which is an ongoing work.

\begin{table}
\centering
\caption{Computation Time for Line Equations with and without SIMD}
\label{tab:time-equations-simd}
\begin{tabular}{lcccc} 
\toprule
          & \begin{tabular}[c]{@{}l@{}}Julia\\(SIMD) \end{tabular} & \begin{tabular}[c]{@{}l@{}}Julia\\(Array) \end{tabular} & \begin{tabular}[c]{@{}l@{}}Python\\(Array) \end{tabular} & MATLAB  \\ 
\midrule
Time (ms) & 0.54                                                   & 5.2                                                     & 12.2                                                     & 1.2      \\
\bottomrule
\end{tabular}
\end{table}

\begin{table}
\centering
\caption{Descriptions of the Six Proposed Workflows}
\label{tab:workflow-descriptions}
\begin{tabular}{@{}ll@{}}
\toprule
\# & Descriptions                                                       \\ \midrule
(1)               & Serial inter-model and intra-model execution                       \\
(2)               & Serial inter-model execution, multi-threaded intra-model execution \\
(3)               & Multi-threaded inter-model execution, serial intra-model execution \\
(4)               & Multi-threaded inter-model and intra-model execution               \\
(5)               & Serial inter-model execution with intra-model SIMD                 \\
(6)               & Multi-threaded inter-model execution with intra-model SIMD         \\ \bottomrule
\end{tabular}
\end{table}

\subsection{Parallel Workflow Comparisons}
The combination of two inter-model execution modes (serial or multi-threaded) 
and three intra-model execution modes (serial, multi-threaded, or SIMD) yields
six possible workflows.
\Cref{tab:workflow-descriptions} describes the six workflows, where Workflow (1) is serial and the other five involve parallelism.
All the workflows are implemented in Julia with the same codebase, ruling out performance discrepancies of programming languages.
The computation times are reported in \Cref{tab:eq-jac-time-workflows}, where the circled number is the workflow number,
and each cell contains the equation time and Jacobian time separated by a slash.
The observations are:
\begin{itemize}
    \item The traditional serial workflow (1) is the slowest, 
    and (2) the inter-model multi-threading alone has limited effect.
    \item Workflow (3) - multi-threading within models - \textbf{speeds up by 5x}, but nested parallelism (4) is as slow as the serial workflow.
    \item Workflow (5) - SIMD within each model with serial model executions - \textbf{has a 10x speedup}.
    Workflow (6) with inter-model multi-threading brings slight improvements to workflow (5).
\end{itemize}

\begin{table}
\centering
\caption{Equation / Jacobian time (ms) for six workflows.}
\label{tab:eq-jac-time-workflows}
\begin{tabular}{l|ll} 
\toprule
\diagbox{Intra-Model}{Inter-Model} & \multicolumn{1}{l}{Serial} & \multicolumn{1}{l}{Multi-threaded}  \\ 
\hline
Serial                             & \circled{1} 5.5 / 21.5                   & \circled{2} 5.3 / 20.9                            \\
Multi-threaded                     & \circled{3} 1.2 / 4.2                    & \circled{4} 5.7 / 20.9                           \\
SIMD                               & \circled{5} 0.6 / 2.8                    & \circled{6} 0.5 / 2.7                             \\
\bottomrule
\end{tabular}
\end{table}

\begin{table}
\centering
\caption{Equation time ($\mu s$) by model in three intra-model workflows.}
\label{tab:time-breakdown}
\begin{tabular}{llllll} 
\toprule
\diagbox{Workflows}{Model } & PQ    & PV    & Slack & Line    & Shunt  \\
\hline
Serial                     & 13.46 & 18.53 & 0.01  & 5172.90 & 1.66   \\
Threaded                                   & 10.81 & 11.10 & 6.50  & 1206.60 & 7.53   \\
SIMD                                       & 17.13 & 7.60  & 0.01  & 520.6   & 1.28   \\
\bottomrule
\end{tabular}
\end{table}

To explain why coarse-grained inter-model multi-threading has limited effects, 
such as in workflow (2) and (6), we break down
the equation computation time by power flow model.
As explained in \ref{sec:parallel-workflow-intro}, the total CPU time of a multi-threaded
parallel program is determined by the bottleneck, which is the Line model in power flow.
This can be verified in \Cref{tab:time-breakdown} that workflows (2)'s and (6)'s time is roughly equal to the time for Line equations. 
Therefore inter-model parallelism is ineffective due to the slow execution of the Line model.

Also, the nested parallelism in workflow (4) has an adverse effect compared with 
workflow (2) due to thread limits.
The processor ends up executing one model in each thread and serially run threads within each model, 
which is similar to workflow (2). 
Even worse, workflow (4) has more overhead than (2) due to the creation and termination of threads,
making it slower than the less-optimized workflow (2).

Therefore, workflow (5) is recommended considering the 
implementation complexity and effectiveness.
Compared with workflow (6), workflow (5) is almost as fast and is single-threaded, which is
far more maintainable. 
Multiple power flow calculation jobs can run in parallel processes
efficiently, as single-threaded jobs will not incur thread
switching overhead.

Finally, \Cref{tab:power-flow-time-synthetic} compares workflow (5) with workflow (1) and the state-of-the-art MATPOWER package \cite{5491276}
to solve Newton power flow for 
four synthetic systems.
MATPOWER's time is the average of five consecutive runs
starting from the second run (to allow for data caching and 
pre-compilation).
Since this work accelerates the equation and Jacobian evaluations,
the linear equation solver time is the same for workflows (1) and (5).
Still, workflow (5) is 8.8\% faster for the 2,000-bus system and 
5.3\% faster for the 70k-bus system.
Workflow (5) has similar performance to MATLAB for the 10k-bus and
the 25k-bus systems, due to MATPOWER's
faster linear equation solver but slower
equation and Jacobian routines.
For large systems, equation and Jacobian time become 
more influential; the proposed workflow (5) thus outperforms 
MATPOWER by 20.1\%.

\begin{table}
\centering
\caption{Synthetic Grid Computation time in milliseconds.}
\label{tab:power-flow-time-synthetic}
\begin{tabular}{lllll} 
\toprule
                       & 2000  & 10k   & 25k   & 70k     \\ 
\hline
Workflow (1) -- Serial  & 41.1  & 245.2 & 632.7 & 2608.9  \\
Workflow (5) -- Serial SIMD   & 37.5  & 225.8 & 624.4 & 2470.5  \\ 
MATPOWER               & 100.0 & 238.0 & 624.0 & 2994.0  \\
\hline
Speed up over Serial   & 8.8\% & 6.6\% & 1.3\% & 5.3\%   \\
Speed up over MATPOWER & 152\% & 3\%   & 0.1\% & 20.1\%  \\
\bottomrule
\end{tabular}
\end{table}

\section{Conclusions}
\label{sec:conclusions}
This letter identifies the most effective parallel workflow 
for power flow calculation by investigates six workflows
using multi-threading and SIMD and vectorization.
The concepts of inter-model and intra-model parallelism are proposed.
Benchmarks using the Synthetic 70k-bus system shows that
(1) coarse-grained inter-model parallelism is ineffective because it 
cannot eliminate the Line model computation bottleneck, 
and (2) serially executed models with SIMD is the most effective approach
owning to its 10x speed up and single-threadedness, which avoids threading
overhead.

Future work involves benchmarking the proposed parallelism workflows with the Fast Decoupled method on heterogeneous computing devices, including CPUs and GPUs.

\ifCLASSOPTIONcaptionsoff
  \newpage
\fi

\bibliographystyle{IEEEtran}
\bibliography{IEEEabrv,papers}

\begin{IEEEbiography}[{\includegraphics[width=1in,height=1.25in,clip,keepaspectratio]{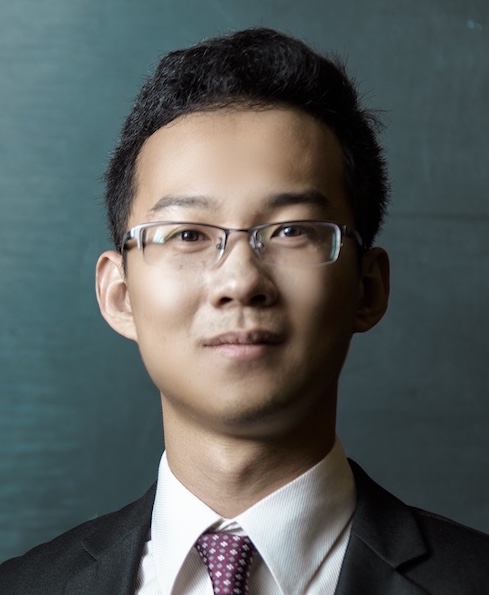}}]{Hantao Cui}
(SM'20) is a research assistant professor with CURENT and the Department of Electrical Engineering and Computer Science, University of Tennessee, Knoxville, where he received his Ph.D. degree in 2018.

He serves as associate editor for the \textsc{Journal of Modern Power Systems and Clean Energy}. He is recognized as an Outstanding Reviewer for 2019 and 2020 by the \textsc{IEEE Transactions on Power Systems} and is author of two Essential Science Indicators (ESI) highly cited papers. His research interests include power system modeling, simulation, and high-performance computing.
\end{IEEEbiography}
\begin{IEEEbiography}[{\includegraphics[width=1in,height=1.25in,clip,keepaspectratio]{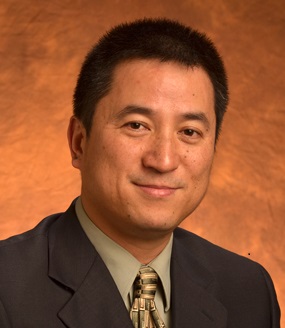}}]{Fangxing (Fran) Li} 
(F'17) received the B.S.E.E. and M.S.E.E. degrees from Southeast University, Nanjing, China, in 1994 and 1997, respectively, and the Ph.D. degree from Virginia Tech, Blacksburg, VA, USA, in 2001. He is currently the James McConnell Professor with the University of Tennessee, Knoxville, TN, USA.

His research interests include deep learning in power systems, renewable energy integration, demand response, power market and power system computing. Since 2020, he has been the Editor-In-Chief for the \textsc{IEEE Open Access Journal of Power and Energy.}
\end{IEEEbiography}

\begin{IEEEbiography}[{\includegraphics[width=1in,height=1.25in,clip,keepaspectratio]{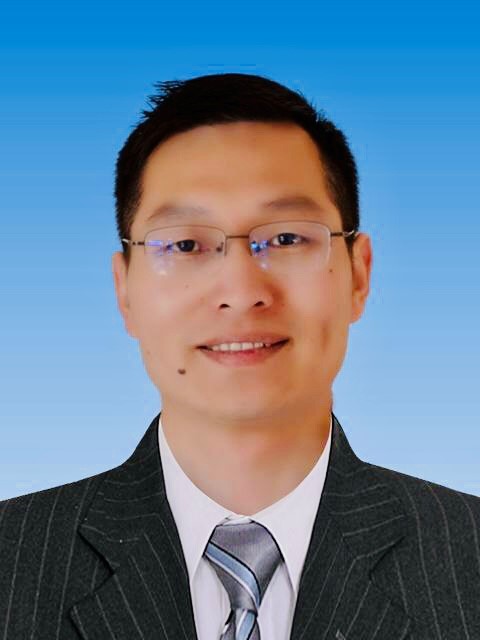}}]{Xin Fang} 
(SM’18) received the Ph.D. degree at the University of Tennessee (UT), Knoxville, TN, USA, in 2016, the M.S. degree from China Electric Power Research Institute, China, in 2012, and the B.S. degree from Huazhong University of Science and Technology, China, in 2009. He is currently with National Renewable Energy Laboratory (NREL). His research interests include power system planning and optimization, electricity market operation considering renewable energy integration, and cyber-physical transmission and distribution modeling and simulation.
\end{IEEEbiography}

\end{document}